\documentclass{book}    

\usepackage{piers}  
\usepackage[]{graphicx}
\usepackage{graphicx}

\usepackage{enumitem}
\usepackage{amsmath}
\usepackage{amssymb}
\usepackage{epsfig}
\usepackage{cite}
\usepackage{color}
\usepackage{balance}
\usepackage{pgfplots}
\usepackage{wrapfig}
\usepackage{lipsum}
\usepackage{hyperref}
\usepackage{gensymb}

\usepackage{nicefrac}       
\usepackage{microtype}      
\usepackage{relsize}
\usepackage{multicol}
\usepackage{multirow}
\usepackage{array}

\begin{document}

\title{Analyzing Air Pollutant Concentrations in New Delhi, India}
\maketitle

\author      {F. M. Lastname}
\affiliation {University}
\address     {}
\city        {Boston}
\postalcode  {}
\country     {USA}
\phone       {345566}    
\fax         {233445}    
\email       {email@email.com}  
\misc        { }  
\nomakeauthor

\author      {F. M. Lastname}
\affiliation {University}
\address     {}
\city        {Boston}
\postalcode  {}
\country     {USA}
\phone       {345566}    
\fax         {233445}    
\email       {email@email.com}  
\misc        { }  
\nomakeauthor

\begin{authors}

{\bf Bugra Alparslan}$^{1}$$^{\dagger}$, 
{\bf Mayank Jain}$^{2,3}$$^{\dagger}$, 
{\bf Jiantao~Wu}$^{2,3}$$^{*}$
{\bf and Soumyabrata Dev}$^{2,3,4}$ \\
\medskip

$^{1}$Middle East Technical University (METU), Ankara, Turkey\\

$^{2}$ADAPT SFI Research Centre, Dublin, Ireland\\

$^{3}$School of Computer Science, University College Dublin, Ireland.

$^{4}$Beijing-Dublin International College, Beijing, China.

$^{*}$ Presenting author and corresponding author\\

$^{\dagger}$ Authors contributed equally

\end{authors}

\begin{paper}

\begin{piersabstract}
Air pollutants have long been known to cause major health problems across humans and all living organisms. Apart from that, they also play a crucial role in temperature inversion situations in the atmospheric layers thereby seriously impacting the radio communications, increased fog levels and decreased visibility. Appreciating the seriousness of these pollutants, this paper attempts to analyze and create a publicly available and easily accessible dataset of seven different pollutants for New Delhi region in India. This analysis and pre-processing is done to assist the researchers who wish to use the dataset for further studies like pollutant forecasting or correlation analysis, thereby promoting the research in the domain.
\end{piersabstract}

\psection{Introduction}
Apart from causing major health problems in humans, increased level of air pollutants in atmosphere have also been noted to be intimately linked with temperature inversion situations in the atmospheric layers. While temperature at a lower atmospheric layer is normally higher than a layer at a higher altitude, inversion of this phenomenon is referred to as the temperature inversion. This inversion causes air pollutants to trap in a lower atmospheric layer and makes the situation of air pollution worse. When combined with the condition of increased relative humidity, chances of creation of fog increases as well which in itself has many serious implications on human health. Low lying inversions (which are directly correlated with anthropogenic emissions of air pollutants) often also changes the profile of atmospheric refraction index causing the microwave beams to entrap in the lower layer impacting the radio communications~\cite{david2016using}.
Consecutively, remote sensing analysts continuously monitor the amount of pollutants in the atmosphere. They are usually performed via satellite images~\cite{kaloni2021impact,manandhar2017correlating}. However, these images suffer from low temporal and low spatial resolution. Therefore, observations recorded from the ground offer a useful alternative~\cite{danesi2021monitoring}.

There are low-cost sensors that continuously record some pollutant concentration levels in the atmosphere (PM$_{10}$, PM$_{2.5}$, CO, Ozone, \textit{etc.}). The National Air Quality Index of India (NAQII)\footnote{\url{https://app.cpcbccr.com/AQI_India/}} has these records of ground-based observations of Delhi, India. The objective of this work is to systematically scrap this NAQII website and archive the data in a user-friendly comma-separated-values (CSV) format and make it publicly available. Additionally, statistical analysis of the pollutant data is conducted to identify trends, seasonality, randomness, and stationarity in the corresponding time series.

\psubsection{Related Work}
Various methods for forecasting air pollutant levels and analyzing their impact on weather and human health have been proposed in the literature. Farah~\textit{et~al.}~\cite{farah2014time} did a detailed analysis on air pollutants in Beirut, Lebanon to determine their persistence, fluctuations and impact on weather. While Chaudhary~\textit{et~al.}~\cite{chaudhary2018time} and Fong~\textit{et~al.}~\cite{fong2020predicting} used recurrent neural networks (RNNs) and long-short-term-memory based networks (LSTMs) respectively to forecast concentration levels of air pollutants, Gul~\&~Khan~\cite{gul2020forecasting} worked on forecasting hazard level of pollutants using LSTMs. A detailed analysis on the short term effects of air pollutants on human health in China was done by Li~\textit{et~al.}~\cite{li2021short} while focusing mainly on pulmonary diseases. Since most widely available datasets have a large number of missing values, Junger~\&~De~Leon~\cite{junger2015imputation} have also proposed a method for imputing the missing values. While a lot of work has been done in the domain of air pollutant analysis, all these datasets were sourced from public websites. Although all of it is in public domain, a lot of time and effort goes in the process of systematic scrapping and pre-processing the data. This inhibits a large number of researchers to quickly try their ideas for analyzing air pollutant datasets, a key factor for speedy progress in research. 

\psection{Dataset}
The national capitol of India, i.e. Delhi, has consistently secured a position in the list of world's top $10$ most polluted cities over the years~\cite{lal1996environmental,worldAQIranking2020}. Hence, the data is scrapped for $2$ years ($2017$ and $2018$) from $2$ stations based in Delhi. These two stations are, namely, the `Anand Vihar' station and the `Punjabi Bagh' station. The readings were taken at a $15$ minute interval for $7$ different type of pollutants, namely, Particulate Matter $10$ ($PM^{10}$), Particulate Matter $2.5$ ($PM^{2.5}$), Nitrogen Dioxide ($N{O}_{2}$), Ammonia ($N{H}_{3}$), Sulphur Dioxide ($S{O}_{2}$), Carbon Monoxide ($CO$), and ground level Ozone (${O}_{3}$). The data is organized month-wise for each pollutant and each station. Since there are a lot of missing values, datasets of all pollutant-month combinations with more than $5\%$ missing values were removed from this study. This led to a total of $58$ monthly entries for different pollutants across the $2$ years and the $2$ stations. Figure~\ref{fig:heatmapMissingValues} shows the heat maps corresponding to the missing value ratio in each month-wise dataset for the considered pollutants.\footnote{The dataset and the code to reproduce results of this paper can be accessed from \url{https://github.com/jain15mayank/air-pollutant-analysis-delhi}}

\begin{figure}[!htb]
    \centering
    \includegraphics[width=0.95\textwidth]{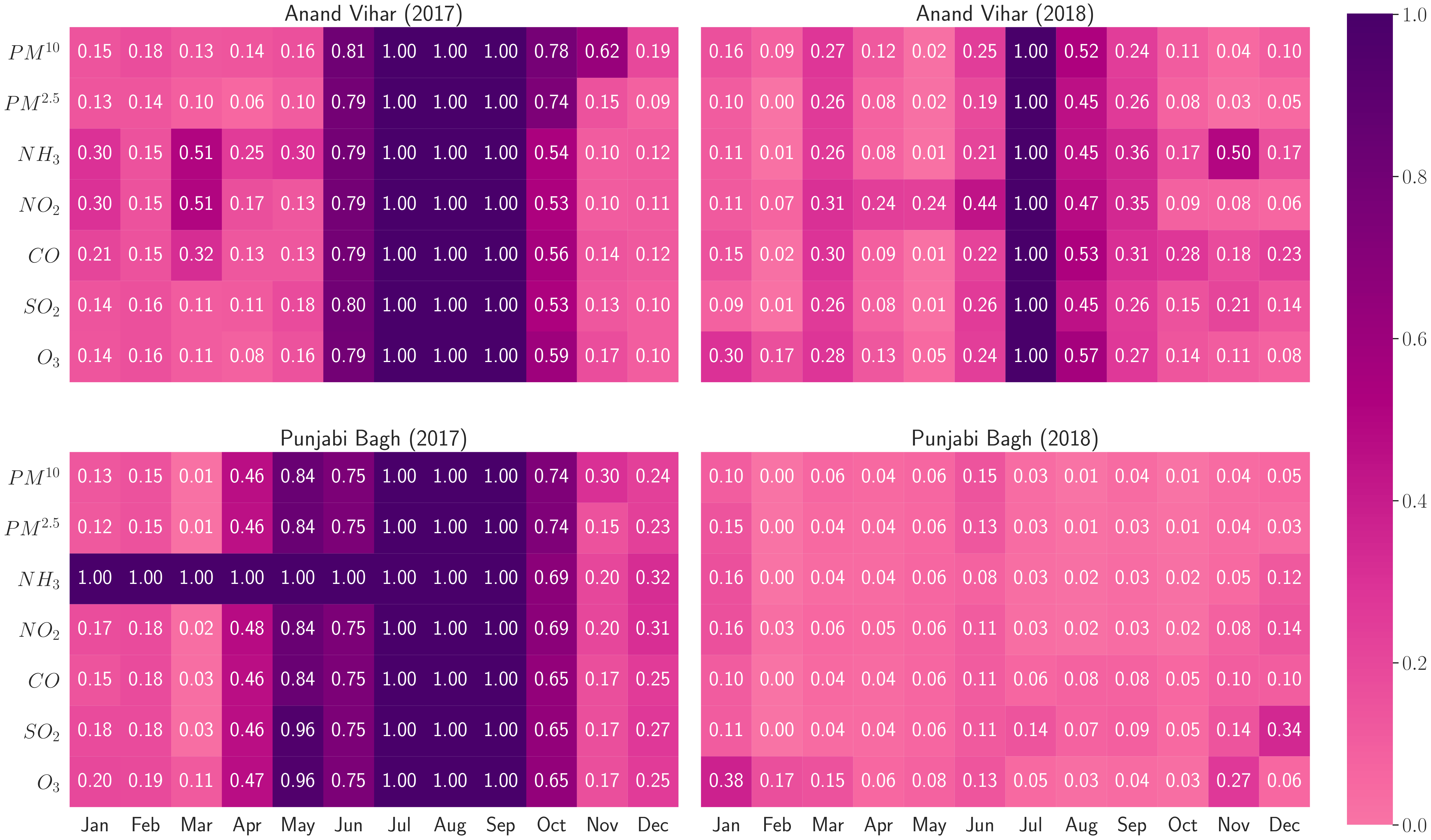}
    \caption{Heat map showing the ratio of missing values to the total number of recordings for different pollutants in different months across the two stations of `Anand Vihar' and `Punjabi Bagh' for $2017$ and $2018$.}
    \label{fig:heatmapMissingValues}
\end{figure}

Figure~\ref{fig:lineCurves} shows the line curves of $4$ sample monthly entries of $4$ different pollutants. These curves are explained as follows:
\begin{enumerate}[label=(\alph*)]
    \item $PM_{2.5}$ readings recorded from `Anand Vihar' station in February, $2018$
    \item $PM_{10}$ readings recorded from `Punjabi Bagh' station in March, $2017$
    \item $CO$ readings from `Anand Vihar' station in May, $2018$
    \item ${SO}_{2}$ readings from `Punjabi Bagh' station in February, $2018$
\end{enumerate}
Rolling mean and rolling standard deviation were also computed with a window of $1$ day and are shown alongside the respective figures.

\begin{figure}[htb]
\centering
\includegraphics[width=0.475\textwidth]{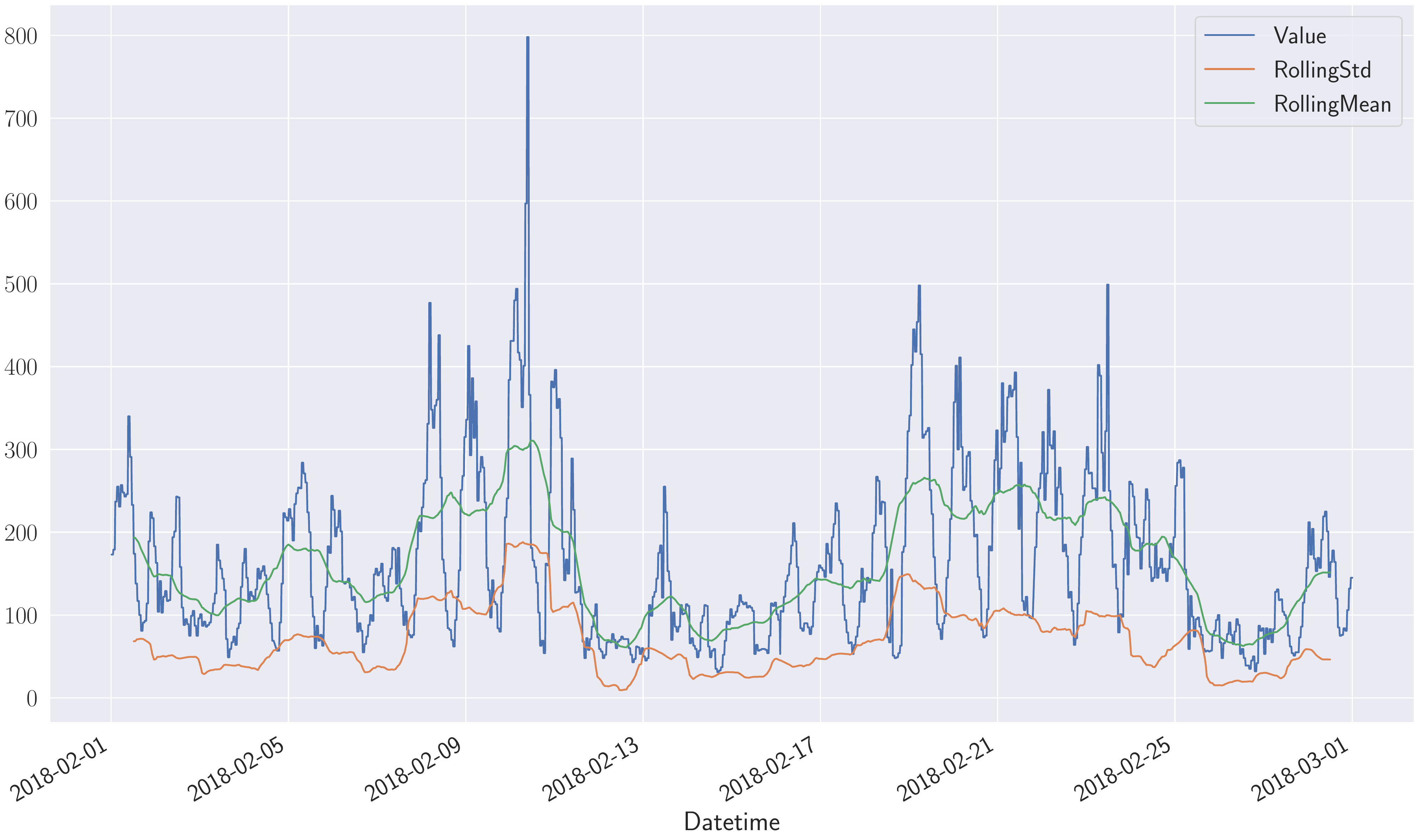}\hfill
\includegraphics[width=0.475\textwidth]{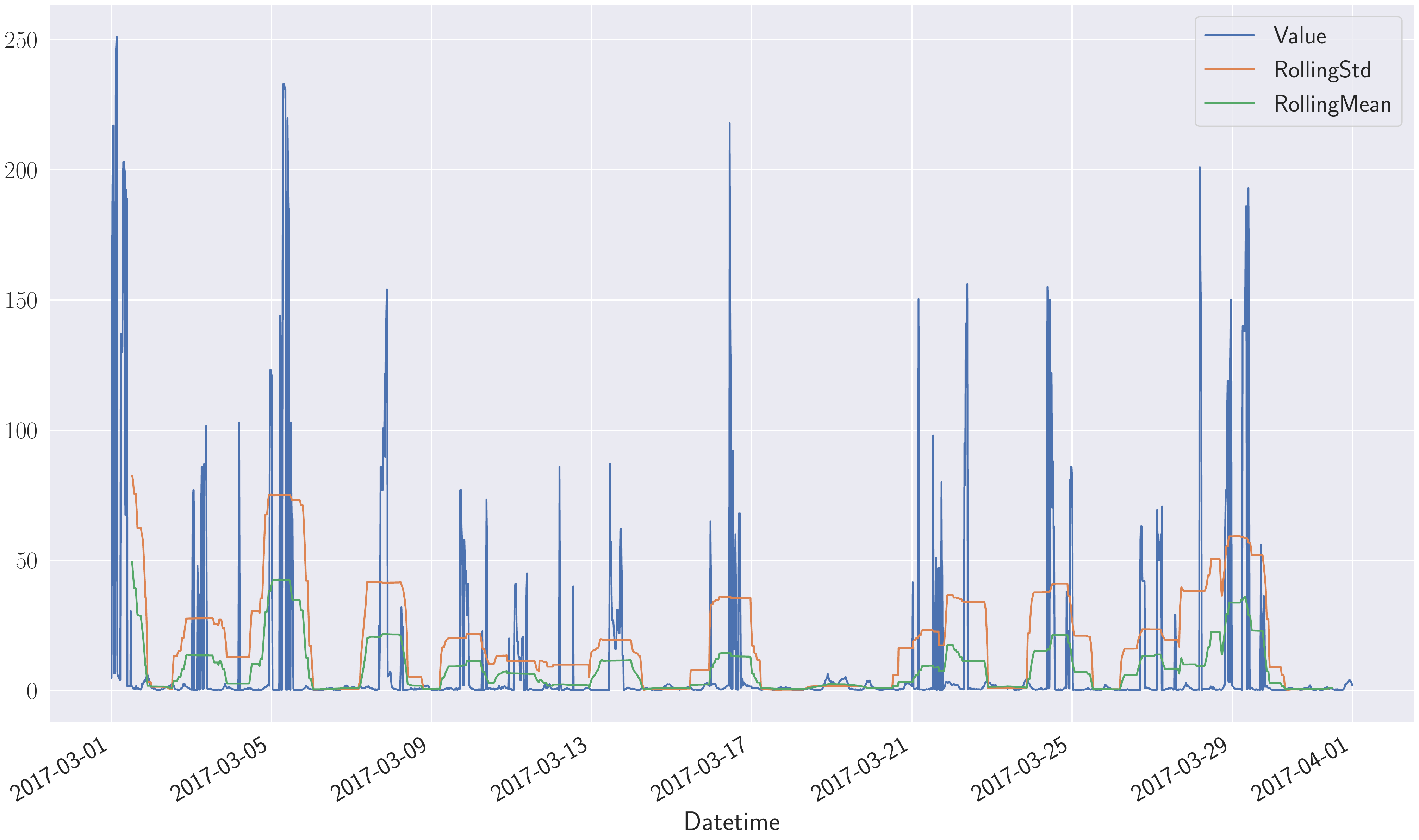}\\
\makebox[0.475\textwidth][c]{(a)}\hfill
\makebox[0.475\textwidth][c]{(b)}\\
\includegraphics[width=0.475\textwidth]{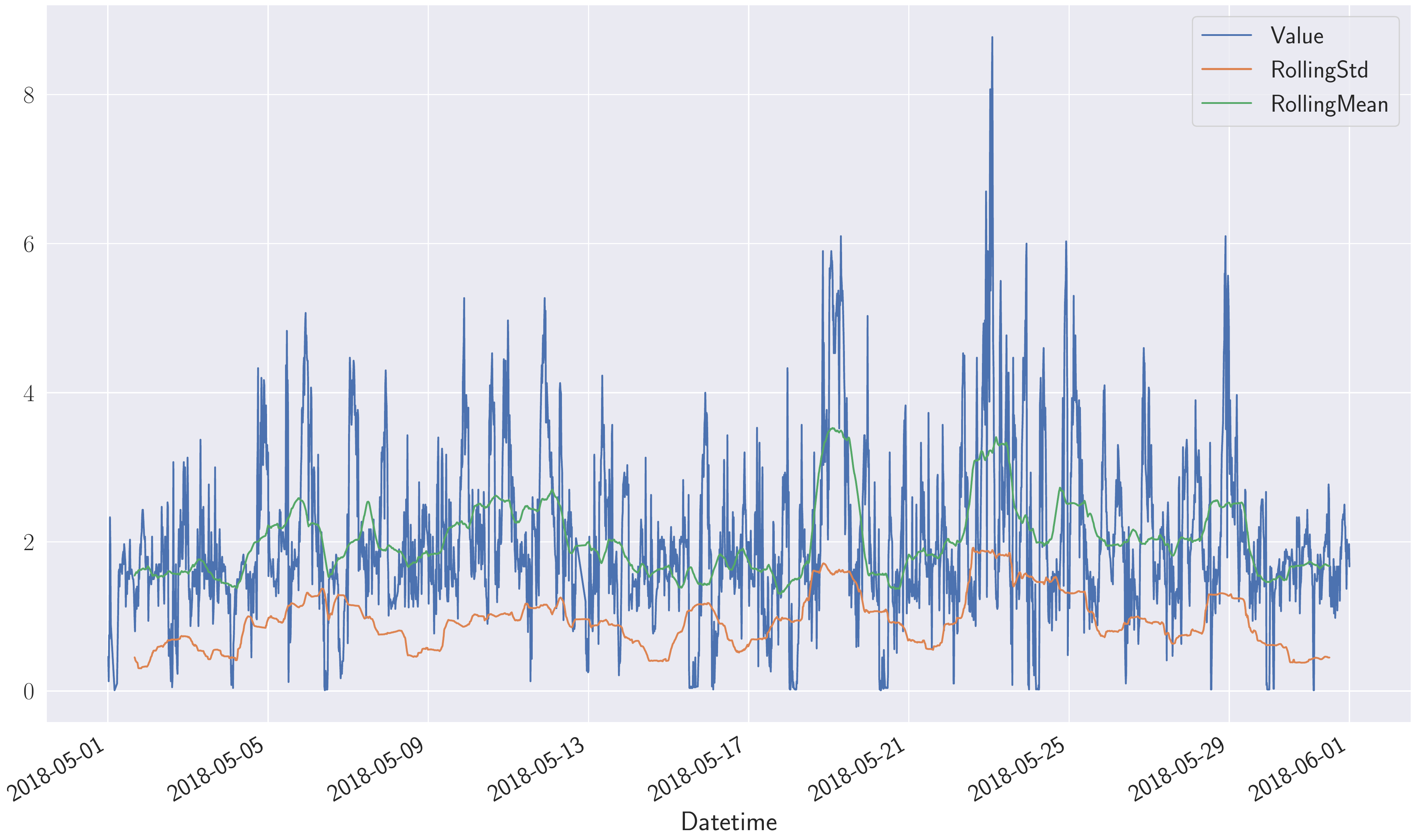}\hfill
\includegraphics[width=0.475\textwidth]{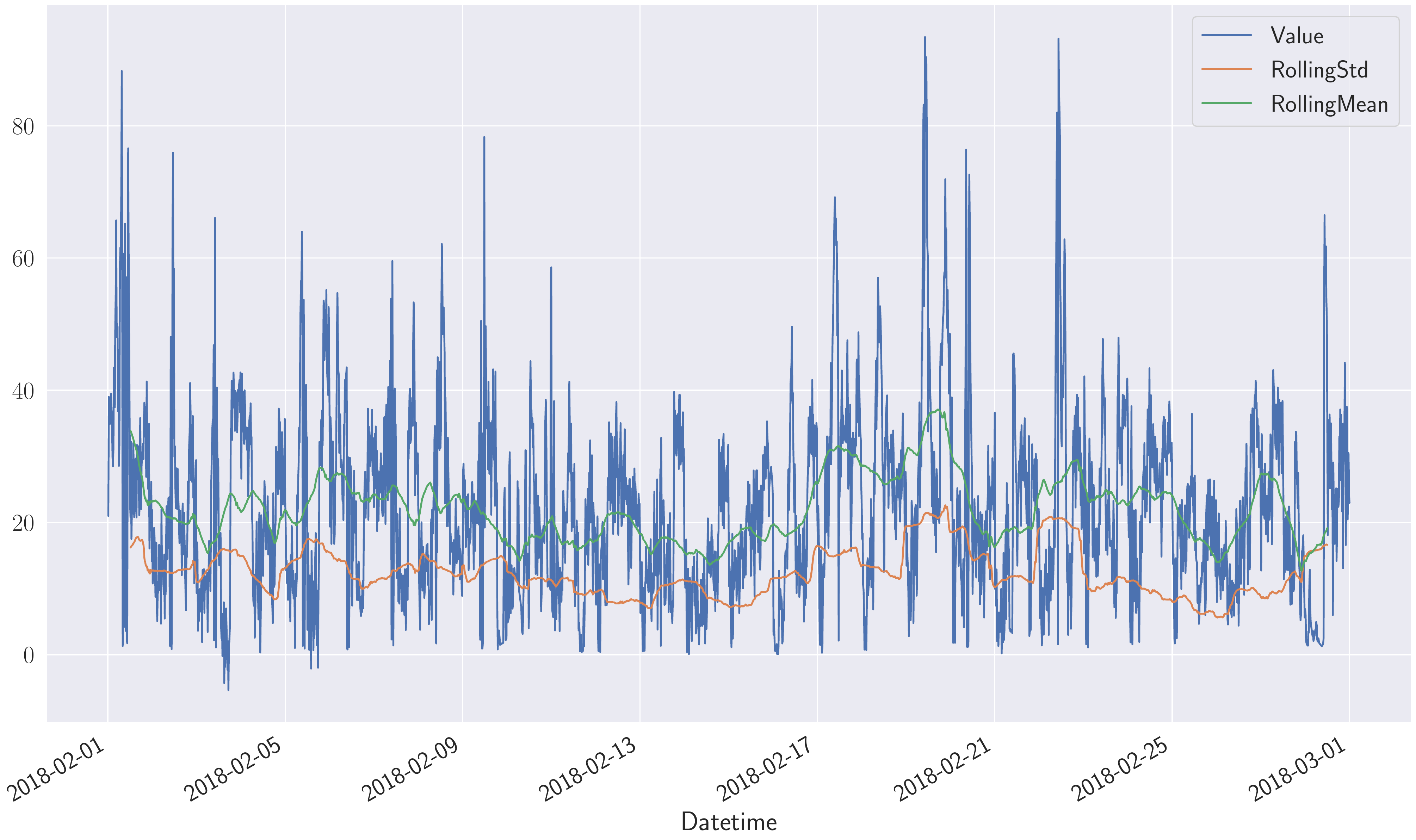}\\
\makebox[0.475\textwidth][c]{(c)}\hfill
\makebox[0.475\textwidth][c]{(d)}
\vspace{-0.1cm}
\caption{In first row, line plots of $PM^{2.5}$ (a) and $PM^{10}$ (b) pollutants captured from `Anand Vihar' station in February $2018$ and `Punjabi Bagh' station in March $2017$ respectively. In second row, line plots of $CO$ (c) and $S{O}_{2}$ (d) pollutants captured from `Anand Vihar' station in May $2018$ and `Punjabi Bagh' station in February $2018$ respectively. All $4$ plots are shown along with their day-wise rolling mean and standard deviation.
}
\label{fig:lineCurves}
\end{figure}

\psection{Data Analysis}
\psubsection{Trend and Seasonality}
A trend is observed when there is an increasing or decreasing slope observed in the time series. The plots shown in figure~\ref{fig:lineCurves} has green lines representing the rolling mean with a window size of $1$ day  or $96$ time steps (as the data is recorded every $15$ minutes). From these plots, it is seen that there is no continuous upward or downward trend in any of the data.

Seasonality occurs when there is a distinct repeated pattern between regular intervals. Although the peak or the lowest points changes from one to another, there are fluctuations and these fluctuations have a period. Running the \textit{Python} code one can see the x-axis values of any point on any graph, and using this it is verified that the time series have a seasonality of approximately $96$ time steps, which makes sense because readings were taken at an interval of $15$ minutes every day.

This seasonality can be better seen from Figure~\ref{fig:decompositionPlots}. These are the decomposition of the data and decomposition is a statistical task in which the time series data is decomposed into several components like trend, seasonality and residuals. The decomposition plots below are shown for the $PM^{10}$ and $PM^{2.5}$ data from the `Anand Vihar' station, captured in November, $2018$.

\begin{figure}[!ht]
    \centering
    \includegraphics[width=0.49\textwidth]{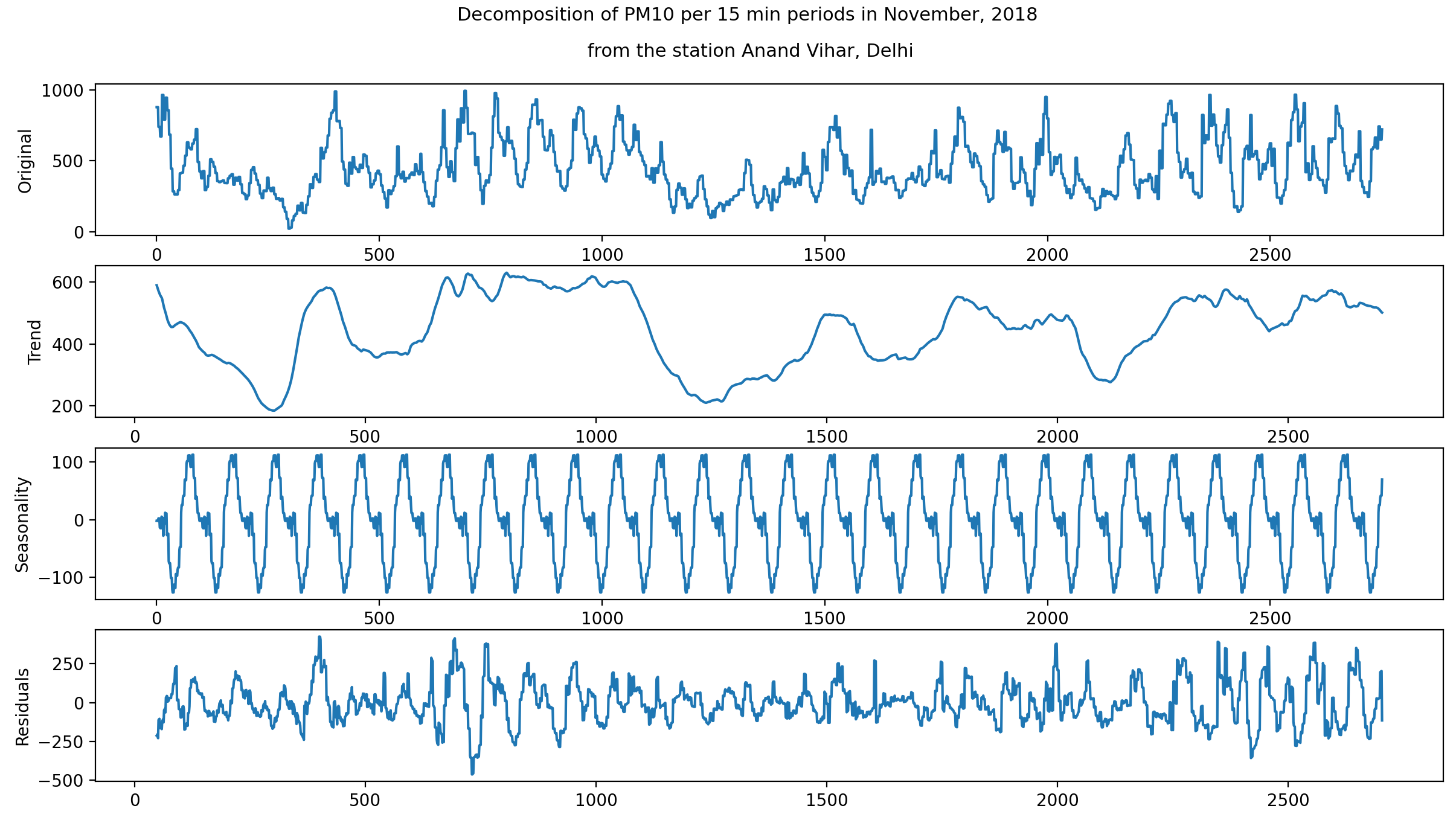}
    \includegraphics[width=0.49\textwidth]{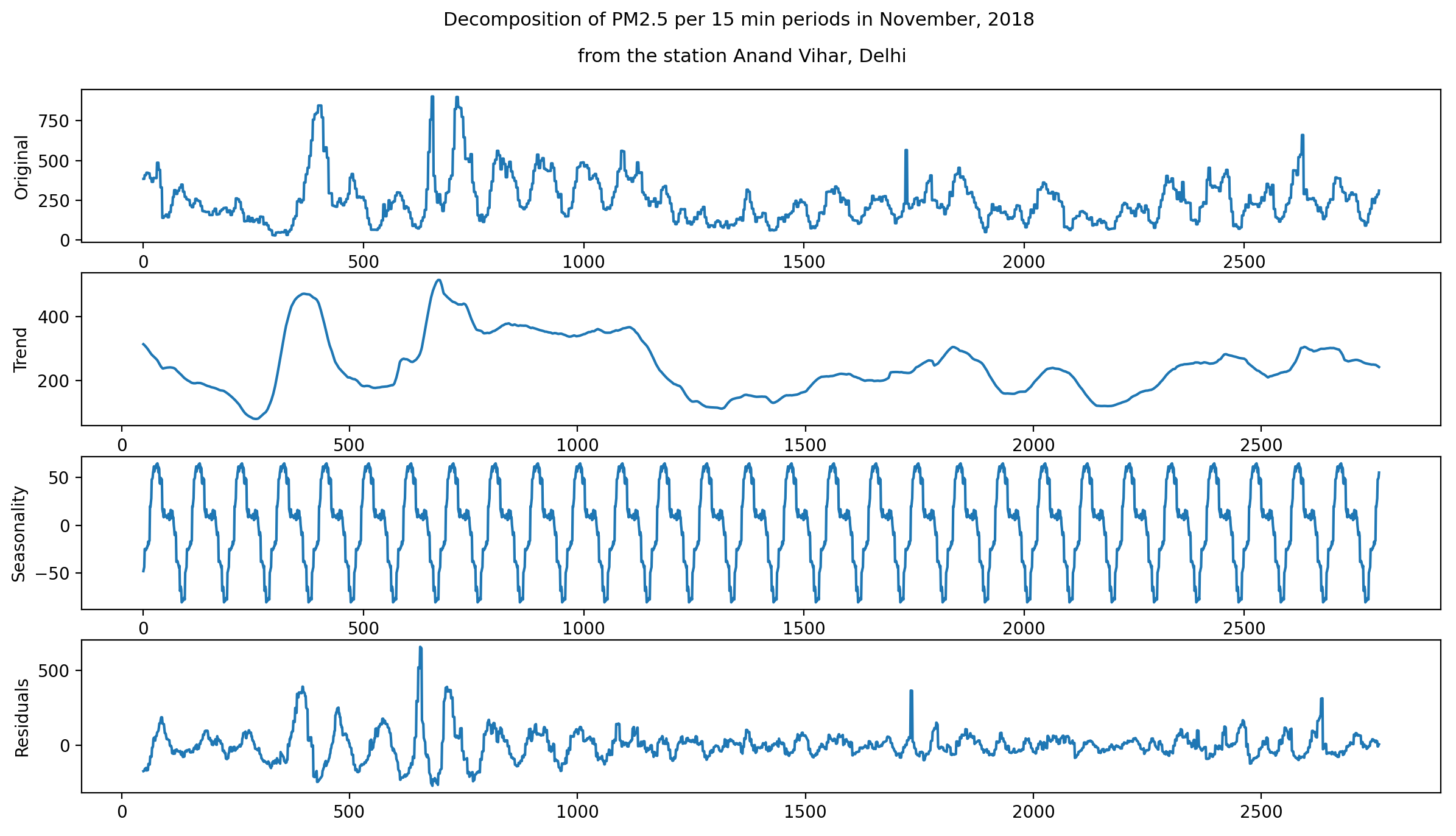}\\
    \makebox[0.49\textwidth][c]{(a)}
    \makebox[0.49\textwidth][c]{(b)}
\caption{Decomposition plots of $PM_{10}$ and $PM_{2.5}$ data from the `Anand Vihar' station, captured in November, $2018$.}
\label{fig:decompositionPlots}
\end{figure}

Figure~\ref{fig:decompositionPlots} help us see that the time series data has no trend but has seasonality. The degree of seasonality is very close to $96$. The last component of the decomposition graphs is residuals. Residuals are what is left over when the seasonal and trend components have been subtracted from the data.

\psubsection{White Noise Time Series}
Next thing to check is whether the time series is white noise or not. A time series is white noise if the variables are independent and identically distributed with a mean of zero. This means that all variables have the same variance ($\sigma^2$) and each value has a zero correlation with all other values in the series. If the time series is white noise, then a model cannot be created to make predictions.

There is a test called Ljung-Box test~\cite{ljung1978measure} to test the lack of fit of a time series model. The Ljung-Box test is defined as:

\begin{center}
\begin{tabular}{l l}
$H_0:$ & The data are independently distributed, random\\
$H_{\alpha}:$ &  The data are not independent, they are related \\
Test Statistic: & Given time series $Y$ of length $n$, the test statistic is defined as: \\
 & $\mathcal{Q} = n(n+2)\mathlarger{\mathlarger{\sum}}_{k=1}^{m}(\frac{\hat{r_k^2}}{n-k}) $ \\
 & where $\hat{r_k}$ is the estimated autocorrelation of the series at lag $k$, \\
 & and $m$ is the number of lags being tested. \\
Significance Level: & $\alpha$ \\
Critical Region: & The Ljung-Box test rejects the null hypothesis\\ 
 & (indicating that the model has significant lack of fit, \\
 & the data is not independent) if \\
 & $\mathcal{Q} > \chi_{1-\alpha, h}^h$ \\
 & where $\chi_{1-\alpha, h}$ is the chi-square distribution table \\ 
 & value with $h$ degrees of freedom and significance level $\alpha$. \\
 \end{tabular}
\end{center}

The degrees of freedom, $h$, is picked according to the result of $min(2m, T/5)$ where $m$ is the period of seasonality and $T$ is the length of time series~\cite{hyndman2018forecasting}. Running the \textit{Python} code, it is verified that time series data used in this paper is not white noise. 

Below is the Table~\ref{tab:ljungbox} which displays the results of the Ljung-Box test for one of the monthly entries that are used in this paper:
 
\begin{table}[!ht]
\renewcommand{\arraystretch}{1.25}
\centering
\begin{tabular}{|l|r|r|r|}
\hline
Lag & p-value & $\mathcal{Q}$         & c-value \\ \hline
1   & 0.000   & 2663.383  & 2.706   \\ \hline
2   & 0.000   & 5075.472  & 4.605   \\ \hline
3   & 0.000   & 7267.983  & 6.251   \\ \hline
4   & 0.000   & 9201.765  & 7.779   \\ \hline
5   & 0.000   & 10887.555 & 9.236   \\ \hline
6   & 0.000   & 12321.927 & 10.645  \\ \hline
7   & 0.000   & 13528.789 & 12.017  \\ \hline
8   & 0.000   & 14524.748 & 13.362  \\ \hline
9   & 0.000   & 15316.793 & 14.684  \\ \hline
\end{tabular}
\caption{\label{tab:ljungbox}Results of Ljung-Box test.}
\end{table}
 
It is seen that the test statistic, $\mathcal{Q}$, is greater than the critical value in each lag. This gives enough evidence to reject the null hypothesis and concludes that the data are not independently distributed, which is the time series is not white noise.

\psubsection{Stationary Time Series}
Time series is stationary if values of the series (mean, variance, autocorrelation) are constant over time, in other words, these values are not a function of time. Forecasting with models like SARIMA or LSTM requires time series to be stationary. The series should be checked if they are stationary, and if not, they should be made stationary.

\psubsubsection{Checking Stationarity}
There are numerous ways to check stationarity, and in this paper it is going to be done by Augmented Dickey-Fuller test and KPSS (Kwiatkowski-Phillips-Schmidt-Shin) test.

\vspace{3mm}
\textbf{Augmented Dickey-Fuller(ADF) Test}\\
The Augmented Dickey-Fuller(ADF) test~\cite{dickey1979distribution,cheung1995lag} is a unit root test. A unit root test determines how strongly a time series is defined by a trend. The ADF test hypothesis are:

\vspace{3mm}
$H_0$: Time series has a unit root, it is non-stationary

$H_1$: Time series does not have a unit root, it is stationary
\vspace{3mm}

To evaluate the test results, a p-value is used. If p-value $> 0.05$, null hypothesis is accepted, which means the data has a unit root and it is not stationary. If p-value $<= 0.05$, null hypothesis is rejected, which means the data does not have a unit root and it is stationary.

Below is a result from an ADF test. This result is taken for one of the monthly entries that are used in this paper.

\begin{center}
    \begin{table}[!ht]
    \renewcommand{\arraystretch}{1.25}
    \centering
    \begin{tabular}{|l|r|}
    \hline
        Test Statistic &  $-6.448849$\\ \hline
        p-value & $1.541445\times{10}^{-08}$\\ \hline
        Number of Lags Used & $28$\\ \hline
        Number of Observations Used & $2846$\\ \hline
        Critical Value ($1\%$) & $-3.43265$\\ \hline
        Critical Value ($5\%$) & $-2.862556$\\ \hline
        Critical Value ($10\%$) & $-2.567311$\\ \hline
    \end{tabular}
    \caption{\label{tab:adftest}Results of ADF test.}
    \end{table}
\end{center}

According to the ADF test, p-value is $1.5\times10^{-8}$. Therefore, the null hypothesis can be rejected, and the data is stationary. The test statistic value is used to determine how likely this rejection can be done: the test statistic value of $-6.4$ is less than the critical value ($1\%$), $-3.4$, this means the null hypothesis can be rejected with a significance level of less than $1\%$.

\vspace{3mm}
\textbf{KPSS (Kwiatkowski-Phillips-Schmidt-Shin) Test}\\
The next test used in this paper is KPSS test. KPSS test is again a unit root test. However, the test hypothesis are different this time:

\vspace{3mm}
$H_0$: Time series does not have a unit root, it is stationary

$H_1$: Time series has a unit root, it is non-stationary
\vspace{3mm}

Table~\ref{tab:kpsstest} is a result of a KPSS test, the data used in this test is the same data used in the ADF test above.

    \begin{table}[htb]
    \renewcommand{\arraystretch}{1.25}
    \centering
    \begin{tabular}{|l|r|}
    \hline
        Test Statistic &  $0.534381$\\ \hline
        p-value & $0.033923$\\ \hline
        Number of Lags Used\quad & $28$\\ \hline
        Critical Value ($10\%$) & $0.347$\\ \hline
        Critical Value ($5\%$) & $0.463$\\ \hline
        Critical Value ($2.5\%$) & $0.574$\\ \hline
        Critical Value ($1\%$) & 0.$739$\\ \hline
    \end{tabular}
    \caption{\label{tab:kpsstest}Results of KPSS test.}
    \end{table}

It is seen that p-value is $0.03$, so the null hypothesis is rejected, the data is not stationary. Test statistic is $0.53$, which is less than the critical value ($1\%$), $0.74$, the null hypothesis is rejected with a significance level of less than $1\%$.

Note that $2$ test give different results. While ADF test describes this data as stationary, KPSS test implies otherwise. The reason for that is there are different types of stationarity, and this two tests check different types of stationarity. As a result, having ADF test result of stationary and KPSS test result of non-stationary means this data is difference stationary, and to make this data stationary, differencing should be applied.

\psubsubsection{Eliminating Stationarity}
Differencing is a method to transform a non-stationary time series into a stationary one. In order to get a differenced data, every value in the dataset should be subtracted from the preceding value. Mathematically it is:
$$z_t = y_t - y_{t-1}$$
where $z_t$ is the differenced data value at time $t$ and $y_t$ is the actual data value at time $t$. Table~\ref{tab:adf_and_kpss} contains the results of ADF and KPSS tests for the differenced time series data.

\begin{table}[ht]
\renewcommand{\arraystretch}{1.25}
\centering
\begin{tabular}{|l|r|l|l|r|}
\cline{1-2} \cline{4-5}
\multicolumn{2}{|l|}{Results of ADF test:}                        & \multirow{9}{*}{} & \multicolumn{2}{l|}{Results of KPSS test:} \\ \cline{1-2} \cline{4-5} 
Test Statistic              & $-16.20271$                      &                   & Test Statistic                & $0.006382$  \\ \cline{1-2} \cline{4-5} 
p-value                     & $4.061694\times{10}^{-29}$                       &                   & p-value                       & $0.1$  \\ \cline{1-2} \cline{4-5} 
Number of Lags Used         & $28$                              &                   & Number of Lags Used           & $28$     \\ \cline{1-2} \cline{4-5} 
Number of Observations Used & $2845$                       &                   & Number of Observations Used   & $-$         \\ \cline{1-2} \cline{4-5} 
Critical Value ($10\%$)       & $-2.567311$                      &                   & Critical Value ($10\%$)         & $0.347$  \\ \cline{1-2} \cline{4-5} 
Critical Value ($5\%$)        & $-2.862556$                      &                   & Critical Value ($5\%$)          & $0.463$  \\ \cline{1-2} \cline{4-5} 
Critical Value ($2.5\%$)      & -                                  &                   & Critical Value ($2.5\%$)        & $0.574$  \\ \cline{1-2} \cline{4-5} 
Critical Value ($1\%$)        & \multicolumn{1}{l|}{$-3.432651$} &                   & Critical Value ($1\%$)          & $0.739$  \\ \cline{1-2} \cline{4-5} 
\end{tabular}
\caption{Results of ADF and KPSS tests for differenced data.}
\label{tab:adf_and_kpss}
\end{table}

Evaluating two tests' results, it is seen that the data is now completely stationary.

\psection{Conclusion and Future Work}
This paper presents a unique dataset of $7$ different pollutant values which was gathered from $2$ different stations in New Delhi over a period of $2$ years (sourced from NQAII website). The dataset was then segregated as month-station-pollutant pairs ($7\times2\times2\times12 = 336$). The paper further performs various statistical tests to comment or process on the trend, seasonality, white noise similarity, and stationarity of the dataset. This analysis and pre-processing is done to assist the researchers who wish to use the dataset for further studies like pollutant forecasting or correlation analysis. In future, the authors would like to extend the dataset by collecting data for more years and more number of stations across the New Delhi area. Furthermore, the plan is to provide a completely pre-processed and cleared version of such datasets in the public domain so that the researchers in the community can make quick and easy use of the data.

\ack
The research in this paper is funded by a research grant from UCD Earth Institute, University College Dublin. The ADAPT Centre  for  Digital  Content  Technology  is  funded under  the  SFI Research Centres Programme (Grant 13/RC/2106\_P2) and is co-funded under the European Regional Development Fund.

\end{paper}

\end{document}